\documentclass{aa}
\usepackage{graphicx}
\usepackage{txfonts}
\usepackage{natbib}
\bibpunct{(}{)}{;}{a}{}{,} \begin{document}

\title{An infrared-submillimeter study of star-forming regions\\
selected by the ISOSS 170\,$\mu$m survey}
\subtitle{}

\author{M.\,Hennemann
\and S.\,M.\,Birkmann
\and O.\,Krause
\and D.\,Lemke}

\offprints{M.\,Hennemann}

\institute{Max-Planck-Institut f\"ur Astronomie (MPIA), K\"onigstuhl 17, D-69117
Heidelberg, Germany\\
\email{hennemann@mpia.de, birkmann, krause, lemke}}

\date{}

\abstract
{Using the ISOPHOT Serendipity Survey (ISOSS) at 170\,$\mu$m a sample of
galactic star-forming regions exhibiting very cold
dust temperatures ($< 20$\,K) and high masses ($> 100$\,M$_\odot$) has been
established.}
{We characterise the star-forming content of five regions that were
selected as potential sites for early stage high-mass star formation.}
{We use SCUBA (JCMT) observations in the submillimeter to identify the dense
condensations of cold gas and dust. Sensitive mid- to far-infared \textit{Spitzer}
observations with IRAC and MIPS allow us to detect associated young stellar objects.
From the long-wavelength
emission we derive dust temperatures and masses for the identified clumps.
A sample of associated mid-infrared sources is investigated using infrared
color-color diagrams and the comparison to a model SED grid to constrain their
evolutionary
stages and derive estimates for additional parameters like the central mass.}
{In every region we identify between one and four submillimeter clumps with
projected sizes between 0.1 and 0.4\,pc.
The dust temperatures range from 11.6 to 21.3\,K and the estimated clump masses
are 2 to 166\,M$_\odot$.
Towards the majority of submillimeter peaks we find point sources in the near-
to mid-infrared. Most are interpreted as low-mass young stellar objects but
we also detect very red sources. They probably represent very young and deeply
embedded protostars that continue to accrete clump material and may reach
higher masses. Several candidate intermediate-mass proto- or
pre-main-sequence stars embedded in the clumps are identified.}
{A subset of four clumps may be massive enough ($> 100$\,M$_\odot$) to form
high-mass stars
and accompanying clusters. The absence of stellar precursors with current masses
in the high-mass regime leave the type of star formation occuring in the
clumps unsettled. We confirm the presence of large fractions of cold
material as derived from large-scale far-infrared measurements which dominates
the emission of most clumps and suggests that the star-forming process will
continue.}


\maketitle

\section{Introduction}

The origin of stars with masses of 10\,M$_\odot$ or more is a
lively debated issue. Only recently approaches to establish an evolutionary
sequence based on observations and theoretical considerations have been presented
\citep{2007prpl.conf..165B,Zinnecker2007}. However, the observational basis
concerning the earliest phases in the formation process of high-mass stars
remains sparse, as many studies target luminous infrared sources revealing
embedded stellar precursors associated with hot as well as high-density gas.
Younger objects that are sometimes addressed as infrared-quiet massive
cores should be best detected using far-infrared and submillimeter dust
continuum surveys as they are assumed to contain large amounts of cold material.
Furthermore, the identification of extinction features against the galactic
mid-infrared background provides a number of candidate objects commonly
referred to as infrared dark clouds (IRDCs).
Only a relatively small number of candidate sources have been investigated in
detail \citep[recently e.g.][]{2007ApJ...656L..85B,2007ApJ...662.1082R,2007A&A...474..883B}
but also results of larger-scale studies suggest that the lifetimes of massive
prestellar cores are very short \citep{2007A&A...476.1243M}.

The thermal emission from cold dust ($< 20$\,K) peaks in the wavelength range
beyond the IRAS 100\,$\mu$m limit. The 170\,$\mu$m ISOPHOT Serendipity
Survey (ISOSS) \citep{1996A&A...315L..64L,1996A&A...315L..71B,2007A&A...466.1205S}
carried out during the ISO mission \citep{1996A&A...315L..27K} provides the first
large-scale survey in this range and is well-suited to search for extremely young
and massive star-forming regions. Selecting compact 170\,$\mu$m sources
(FWHM $<$ 3.5\arcmin)
 that are
associated with an IRAS 100\,$\mu$m point source \citep{1988iras....1.....B} and
molecular gas emission, a sample of more than 50 massive candidate sources has
been identified \citep[see][and references therein]{2003PhDT.........3K,2004BaltA..13..407K}.
The large-scale average dust temperatures derived from the flux
ratios at 170 and 100\,$\mu$m are about 18\,K or lower and the resulting
mass-luminosity ratios are M/L\,$\sim$ 0.6\,M$_\odot$/L$_\odot$, implying early
stages in the process of star formation
  \citep[cf.][]{2002ApJ...566..931S} .
In order to explore the physical
conditions in this unique sample of cold and massive star-forming regions and to
search for possible high-mass prestellar cores, a multi-wavelength follow-up
survey has been launched. The presence
of very young massive clumps and cores has been confirmed for the regions
ISOSS J20298+3559 \citep{2003A&A...398.1007K}, ISOSS J18364-0221
\citep{2006ApJ...637..380B} and
ISOSS J23053+5953 \citep{2007A&A...474..883B}. In this
paper we report on observations of four more ISOSS star-forming regions
and extend the study of ISOSS J20298+3559.

Our continuum observations with high spatial resolution covering the
submillimeter to near-infrared wavelength range allow us to characterize
the detailed star-forming content. In the submillimeter we trace
the thermal dust emission and identify compact condensations. To constrain their
evolutionary stage we have conducted mid- and far-infrared observations using
\textit{Spitzer}. The
long-wavelength data in the far-infrared and submillimeter is used to estimate
the dust temperatures and total masses of the clumps. The sensitive
mid-infrared data also reveal very young associated stellar objects.
Additional deep near-infrared images provide a good assessment of source
confusion. For the detected young stellar objects
we perform an evolutionary classification from their infrared colors. In the
Discussion section we further characterize a subset using radiative transfer
models.

\begin{table*}
\begin{minipage}[h]{\hsize}
\caption{Observed ISOSS star-forming regions.}
\label{tab_regions}
\centering
\renewcommand{\footnoterule}{}  \begin{tabular}{c c c c c c}
\hline\hline
Region & Assoc. IRAS & RA\,(J2000) & DEC\,(J2000)& Distance\footnote{
Kinematic distances obtained using the galactic rotation model of \citet{1993A&A...275...67B}
and identification of kinematically associated molecular cloud complexes \citep{2003PhDT.........3K}.}
& T$_{\rm FIR\,colour}$\footnote{T$_{\rm FIR\,color}$ is derived from the
far-infrared flux ratios at 100\,$\mu$m (IRAS) and 170\,$\mu$m (ISOSS).} \\
ISOSS\dots & point source & & & in kpc & in K \\
\hline
J19357+1950 & 19335+1944 & 19:35:45.9 & +19:50:58 & 4.0 & 17.5\\
J19486+2556 & 19465+2549 & 19:48:36.8 & +25:56:55 & 2.9 & 18.0\\
J20153+3453 & 20134+3444 & 20:15:20.9 & +34:53:53 & 2.0 & 19.0\\
J20298+3559 & 20278+3549 & 20:29:48.3 & +35:59:24 & 1.8 & 16.0\\
J22478+6357 & 22460+6341 & 22:47:54.1 & +63:57:11 & 4.1 & 15.4\\
\hline
\end{tabular}
\end{minipage}
\end{table*}

\section{Observations, Data Reduction and Analysis}

\subsection{Submillimeter continuum emission}
Submillimeter continuum jiggle maps at 450\,$\mu$m and 850\,$\mu$m were obtained with
SCUBA \citep{1999MNRAS.303..659H} at the James Clerk Maxwell Telescope (JCMT) in
July 2001 and May 2003 under good atmospheric transmission conditions
($\tau_{850\,\mu\mathrm{m}} \lesssim 0.2$). Reduction was done using the
ORAC-DR \citep{1999ASPC..172..171J} and SURF \citep{1998ASPC..145..216J} software.
Photometric calibration is based on maps
of Uranus acquired shortly before or after the observations. Further analysis
was done using the MIRIAD \citep{sault-1995-77} software package and followed
the procedure described in \citet{2001ApJS..134..115S}: To account for the deviations
of the JCMT beam from a single Gaussian we used the Uranus maps to construct
symmetric beam models and deconvolve the maps of the target regions. The derived
beam sizes are 7.8 -- 8.8$\arcsec$ at 450\,$\mu$m and 14.7 -- 15.1$\arcsec$ at
850\,$\mu$m. The maps are then restored using a gaussian of 8$\arcsec$ and 14$\arcsec$,
respectively. By fitting
gaussian components to the restored maps we first derive submillimeter continuum
fluxes and deconvolved source sizes at 450\,$\mu$m. We then use
these sizes convolved with the 14$\arcsec$ gaussian
to
extract 850\,$\mu$m fluxes in order to include emission from the same regions only.
The photometric accuracy is estimated to be 30\% at 450\,$\mu$m and 20\% at 850\,$\mu$m and
the pointing uncertainty of the
submillimeter maps is \mbox{2$\arcsec$ rms}.

\subsection{Near-infrared emission}
Near-infrared images in J, H and Ks were taken with the Calar
Alto 3.5\,m telescope using the two prime focus wide field cameras Omega2000
\citep{2003SPIE.4841..343B} and OmegaPrime \citep{1998SPIE.3354..825B}.
Omega2000 features a field of view (FOV) of $15.4\times 15.4\,\mathrm{arcmin^2}$
with a pixel scale of $0.4496\arcsec\,\mathrm{pix^{-1}}$, while the FOV for
OmegaPrime is $6.8 \times 6.8\,\mathrm{arcmin^2}$ with a pixel scale of $\sim
0.4\arcsec\,\mathrm{pix^{-1}}$. The exposure time in the broad band filters was
20 minutes each. The exposures
were dithered on source to allow for sky subtraction. The reduction and photometry
was done using IRAF \citep{1993ASPC...52..173T} and
GAIA\footnote{http://star-www.dur.ac.uk/\~{}pdraper/gaia/gaia.html} and the
photometric calibration for J, H, and Ks is based on the 2MASS point source catalog.

\subsection{Mid- and far-infrared emission}
The \textit{Spitzer} \citep{2004ApJS..154....1W} observations include IRAC
\citep{2004ApJS..154...10F} imaging in all four photometric bands, MIPS
\citep{2004ApJS..154...25R} imaging at 24\,$\mu$m and 70\,$\mu$m and the MIPS
spectral energy distribution (SED) mode.
For the imaging observations the basic flux calibrated data (BCD) of the
\textit{Spitzer Science Center} (SSC) pipeline was used for further data reduction
and analysis. The calibration uncertainties of the data are about 2\% for IRAC
\citep{2005PASP..117..978R}, 4\% for MIPS 24\,$\mu$m \citep{2007PASP..119..994E}
and 10\% for MIPS 70\,$\mu$m \citep{2007PASP..119.1019G}. Cosmetic
corrections and astrometric refinement was performed with the MOPEX software
\citep{2005PASP..117.1113M} and final images were combined using scripts
in IRAF. Aperture photometry and PSF fitting was done with the
aperture corrections given on the SSC
website\footnote{http://ssc.spitzer.caltech.edu}. The MIPS SED
mode calibration is based on a spectrum of $\alpha$ Boo
\citep{2005ApJ...631.1170L} and the measured MIPS 70\,$\mu$m fluxes. The resulting
photometric accuracy is estimated to be 5\% (IRAC), 10\% (MIPS 24), and 20\% (MIPS
70 and SED).

\section{Observational results: Submillimeter emission morphology
and associated mid-infrared sources}

The five ISOSS regions are listed in Table \ref{tab_regions} and displayed in
Figs.~\ref{fig:19357_maps} to \ref{fig:22478_maps}. In the following
we describe their individual submillimeter morphologies and point out the
associated 24\,$\mu$m sources. On the IRAC and near-infrared maps those sources
for which singular counterparts can be identified at shorter wavelengths are
marked.

\subsection{ISOSS J19357+1950}

\begin{figure*}
\centering
\includegraphics[width=18cm]{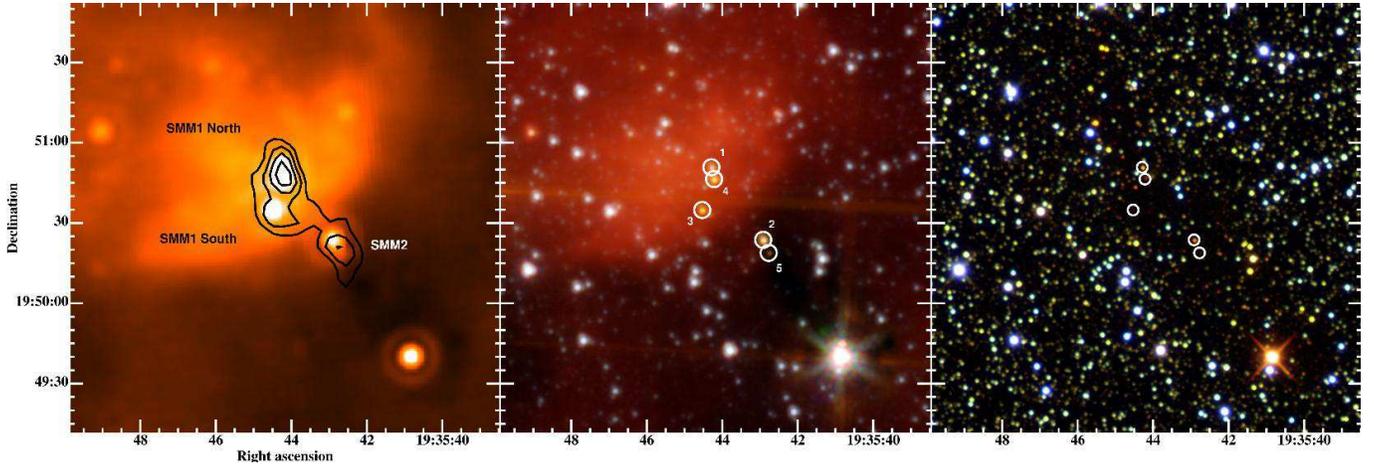}
\caption{
Observations of the region ISOSS J19357+1950. The left panel
shows the 24\,$\mu$m map and the overlaid contours show the SCUBA 450\,$\mu$m
emission. The mid panel shows a color composite of the IRAC images. The right
panel is
a JHKs color composite using OmegaPrime and Omega2000 observations.
The circles surround the sources investigated in the Analysis section.
} \label{fig:19357_maps}
\end{figure*}

In this region three submillimeter emission components are resolved at 450\,$\mu$m,
two adjacent parts (SMM1 North and South) and a south-western part (SMM2)
(Fig.~\ref{fig:19357_maps}). They have deconvolved
FWHM sizes of about 19$\arcsec$, 16$\arcsec$ and 20$\arcsec$ which correspond
to $0.31-0.39$\,pc. An arc-shaped
extended emission feature that stretches from South-east to North-west is
visible at 24\,$\mu$m. Two 24\,$\mu$m point sources are detected associated with
SMM1 North, for SMM1 South and SMM2 one and two sources are present respectively.
Both SMM1 North and South are also coincident with emission
at 70\,$\mu$m but are not resolved individually. At the location of SMM2
no distinct 70\,$\mu$m feature is detected.

\subsection{ISOSS J19486+2556}

\begin{figure*}
\centering
\includegraphics[width=18cm]{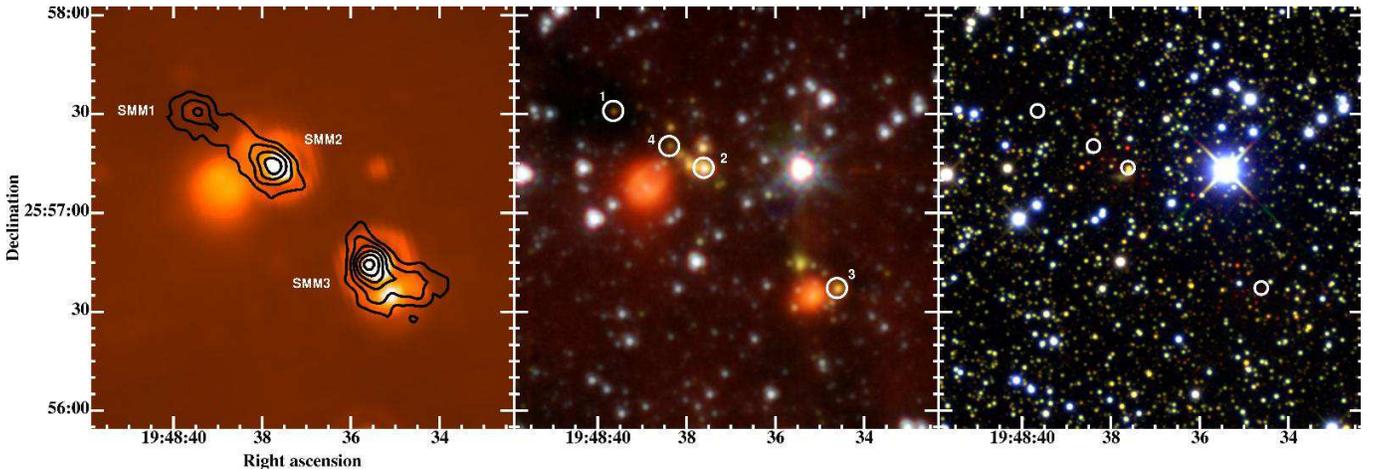}
\caption{
Observations of the region ISOSS J19486+2556.
Same arrangement and symbols as in Fig.~\ref{fig:19357_maps}.
} \label{fig:19486_maps}
\end{figure*}

Three submillimeter clumps are detected in this region: They are located along a
chain
from north-east to south-west and termed SMM1 to SMM3 (Fig.~\ref{fig:19486_maps}).
All three appear compact in the
submillimeter: The deconvolved FWHM diameters are about 16$\arcsec$ (SMM1), 10$\arcsec$ (SMM2) and
15$\arcsec$ (SMM3) at 450\,$\mu$m. The corresponding length scales are $0.14-0.22$\,pc.
Both SMM2 and SMM3 are associated with bright 24 and 70\,$\mu$m sources. In the case
of SMM2, several objects are detected at 24\,$\mu$m and the brightest one
coincides with the submillimeter peak as well as with the 70\,$\mu$m source.
For SMM3, one source at 24 and 70\,$\mu$m is located at the submillimeter peak.
In addition, two areas of extended 24 and 70\,$\mu$m emission are found, one
in the south-east of SMM2 and a second one towards the southern rim of SMM3.
The first one is not associated with submillimeter emission. Only a faint source
at 24\,$\mu$m is detected towards the center of SMM1 and it is not
associated with compact 70\,$\mu$m emission.

\subsection{ISOSS J20153+3453}

\begin{figure*}
\centering
\includegraphics[width=18cm]{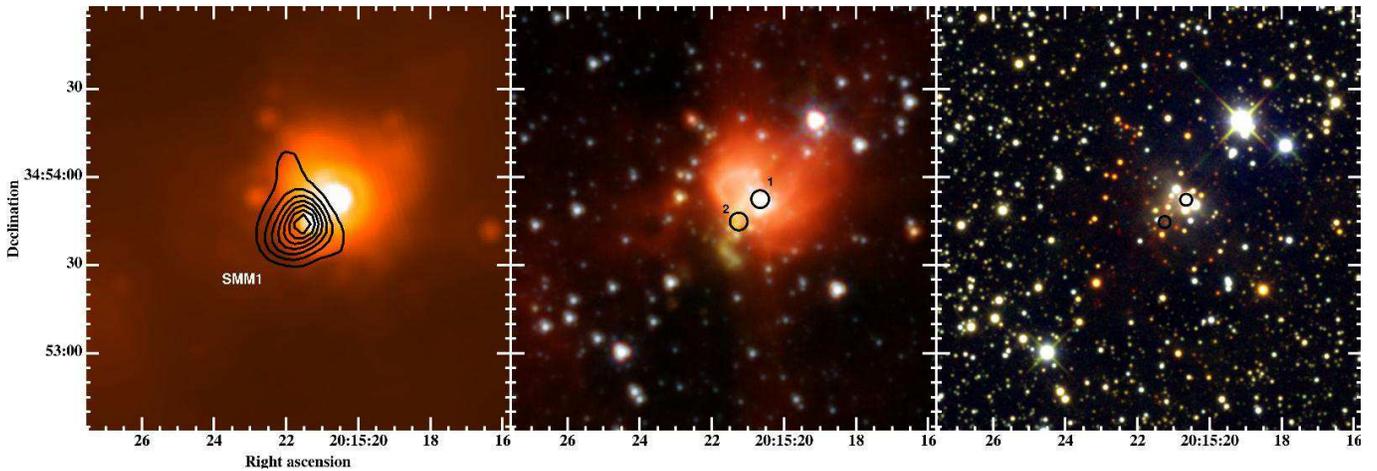}
\caption{
Observations of the region ISOSS J20153+3453.
Same arrangement and symbols as in Fig.~\ref{fig:19357_maps}.
} \label{fig:20153_maps}
\end{figure*}

This region contains a single submillimeter clump clearly detected in both SCUBA
bands (Fig.~\ref{fig:20153_maps}). The deconvolved 450\,$\mu$m FWHM extension is about
$18\arcsec\times14\arcsec$ i.e. the
projected diameter of the clump is around 0.16\,pc.
In the 24\,$\mu$m band two point-like sources surrounded by
extended emission are detected towards the north-western limb of the submillimeter
clump. The brighter one is located among a cluster of sources
seen at shorter wavelengths. At 70\,$\mu$m none of these are resolved individually,
they blend into a single slightly extended source.
The submillimeter emission peak is offset from the
brighter point source by 11$\arcsec$.

\subsection{ISOSS J20298+3559}

\begin{figure*}
\centering
\includegraphics[width=18cm]{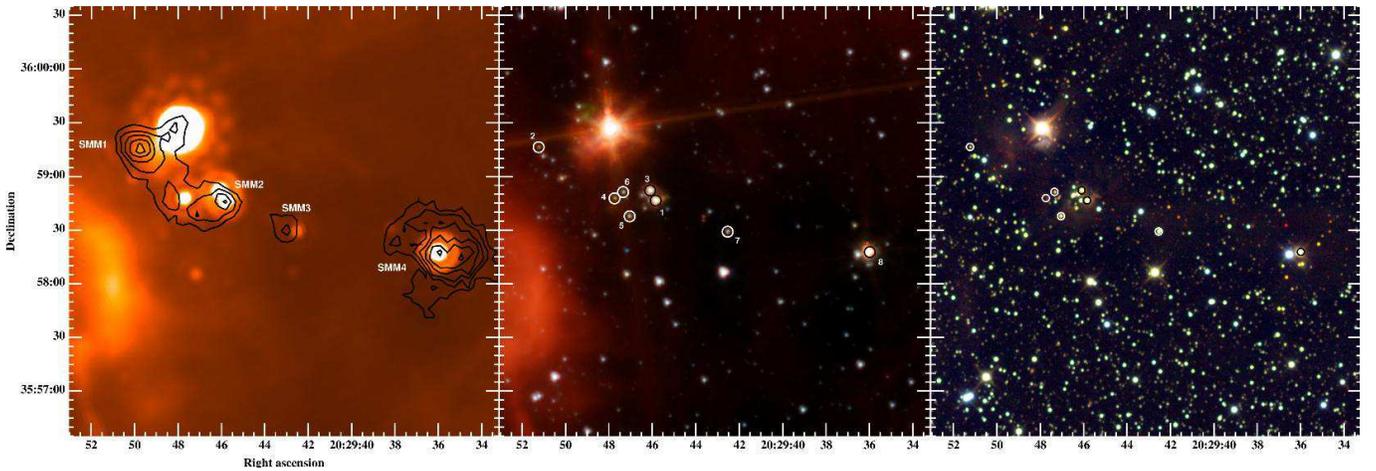}
\caption{
Observations of the region ISOSS J20298+3559. The left panel
shows the 24\,$\mu$m map and the overlaid contours show the SCUBA 450\,$\mu$m
emission (for SMM3 the 850\,$\mu$m emission is shown). The mid panel shows a
color composite of the IRAC images. The right panel is
a JHKs color composite using OmegaPrime and Omega2000 observations.
The circles surround the sources investigated in the Analysis section.
} \label{fig:20298_maps}
\end{figure*}

This region has been studied in detail in \citet{2003A&A...398.1007K}. Four
submillimeter emission peaks are found in this region: Two of them are joined by
extended emission (SMM1 and SMM2) while SMM3 and SMM4 are offset to the west
(Fig.~\ref{fig:20298_maps}).
SMM4 has not been discussed in \citet{2003A&A...398.1007K}. The sizes of the clumps
SMM1 and SMM3 are approximately 0.14 and 0.17\,pc, SMM2 is unresolved ($< 0.07$\,pc)
and SMM4 is more extended: The deconvolved FWHM diameter is
roughly 39$\arcsec$ corresponding to 0.34\,pc. SMM1, SMM2 and SMM4 are associated
with emission
at 70\,$\mu$m. SMM3 was not covered by our 70\,$\mu$m maps. Two 24\,$\mu$m
point sources are detected towards SMM2 and one towards SMM4. SMM1 and SMM3 have
faint counterparts at 24\,$\mu$m.

\subsection{ISOSS J22478+6357}

\begin{figure*}
\centering
\includegraphics[width=18cm]{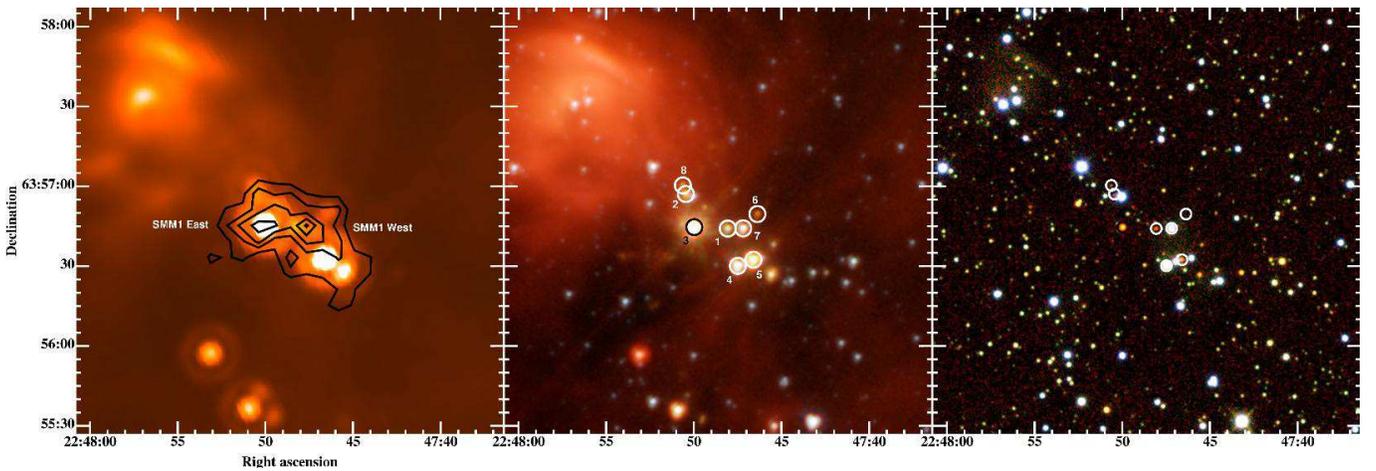}
\caption{Observations of the region ISOSS J22478+6357.
Same arrangement and symbols as in Fig.~\ref{fig:19357_maps}.
} \label{fig:22478_maps}
\end{figure*}

The submillimeter emission in this region traces an elongated clump (SMM1) that
is resolved in an eastern and a western component at 450\,$\mu$m
(Fig.~\ref{fig:22478_maps}). The deconvolved 450\,$\mu$m FWHM extensions of these are
$\sim$\,7$\arcsec$ and $\sim$\,12$\arcsec$, corresponding to roughly 0.14
and 0.24\,pc. At 24\,$\mu$m, several point sources are associated. The brightest
are located at the eastern and to the southern limb of the western submillimeter
peaks. The 70\,$\mu$m map reveals emission that coincides with the submillimeter
peaks and with the 24\,$\mu$m sources.

\section{Analysis}

\begin{figure*}
\centering
\includegraphics[width=18cm]{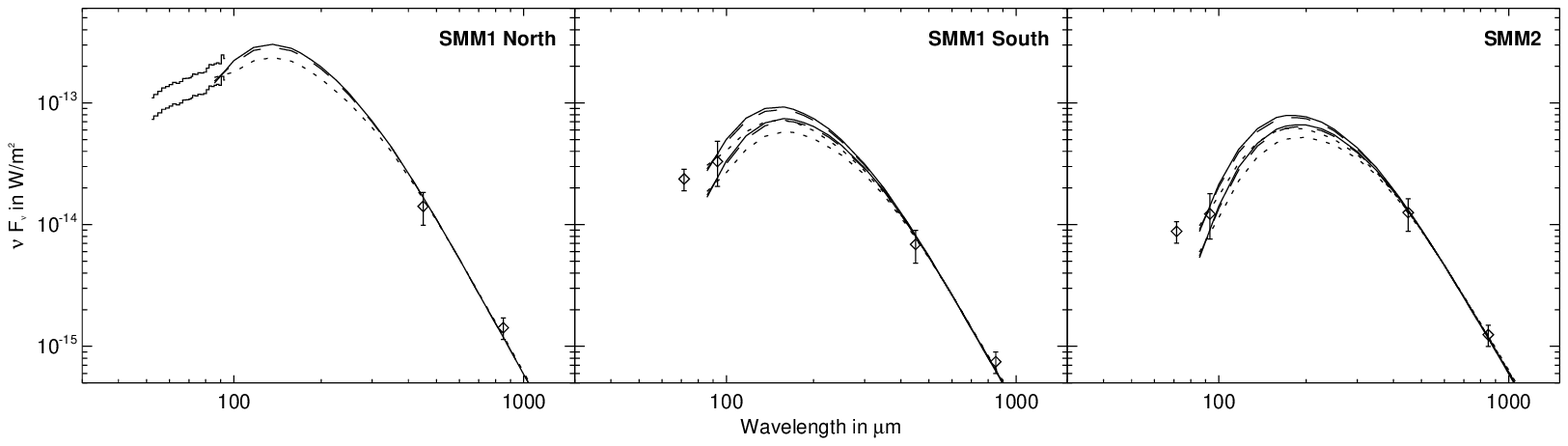}
\caption{
Spectral energy distributions of the submillimeter sources in the
region ISOSS J19357+2556.
The diamonds represent photometric data obtained with MIPS (70\,$\mu$m) and
SCUBA (450 and 850\,$\mu$m) and extrapolated data at 93\,$\mu$m for SMM1 South
and SMM2.
Histogram-like bars show the error range of the MIPS SED spectrophotometry
(53-93\,$\mu$m) for SMM1 North.
The lines show modified blackbody fits for $\lambda \ge 93\,\mu$m using
different dust opacities (solid: no ice mantles, dotted: thick ice mantles,
dashed: thin ice mantles).
For SMM1 South and SMM2 two sets of curves are shown for spectral slopes
of 0.3 and 2.0 in $70 \le \lambda \le 93\,\mu$m.
}\label{fig:19357_seds}
\end{figure*}

\begin{figure*}
\centering
\includegraphics[width=18cm]{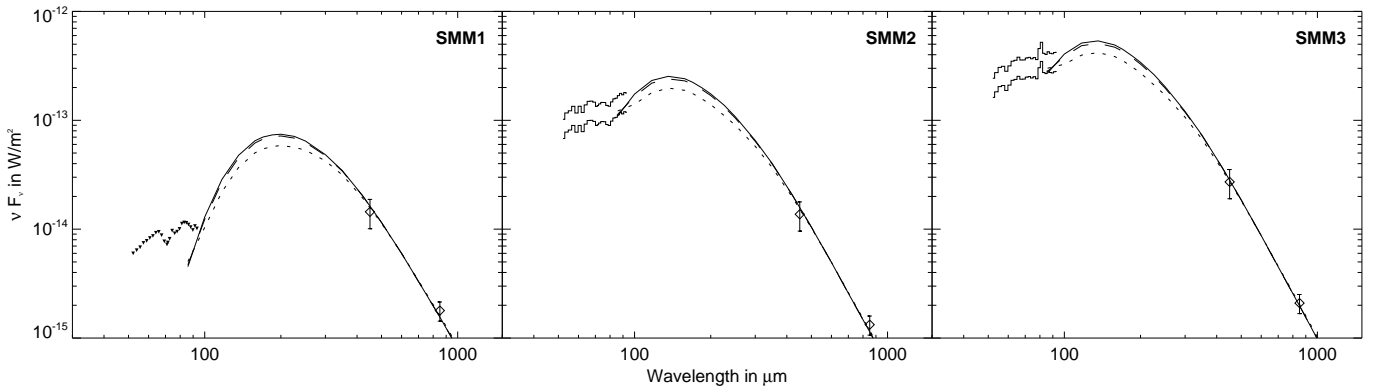}
\caption{
Spectral energy distributions of the submillimeter sources in the
region ISOSS J19486+2556.
The diamonds represent photometric data obtained with
SCUBA (450 and 850\,$\mu$m).
Histogram-like bars show the error range of the MIPS SED spectrophotometry
(53-93\,$\mu$m), for SMM1 upper limits are plotted with triangles.
The lines show modified blackbody fits for $\lambda \ge 93\,\mu$m using
different dust opacities (solid: no ice mantles, dotted: thick ice mantles,
dashed: thin ice mantles).
}\label{fig:19486_seds}
\end{figure*}

\begin{figure}
\centering
\resizebox{6.665cm}{!}{\includegraphics{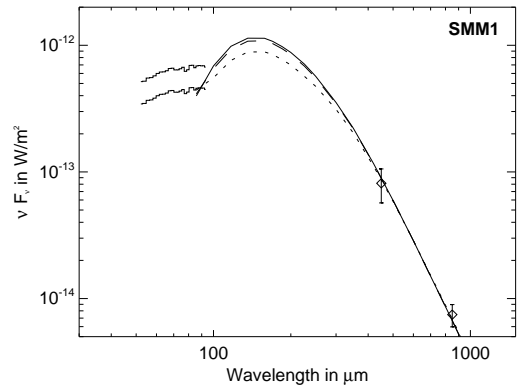}}
\caption{
Spectral energy distribution of the submillimeter source in the
region ISOSS J20153+3453.
The diamonds represent photometric data obtained with
SCUBA (450 and 850\,$\mu$m).
Histogram-like bars show the error range of the MIPS SED spectrophotometry
(53-93\,$\mu$m).
The lines show modified blackbody fits for $\lambda \ge 93\,\mu$m using
different dust opacities (solid: no ice mantles, dotted: thick ice mantles,
dashed: thin ice mantles).
}\label{fig:20153_sed}
\end{figure}

\begin{figure*}
\centering
\includegraphics[width=18cm]{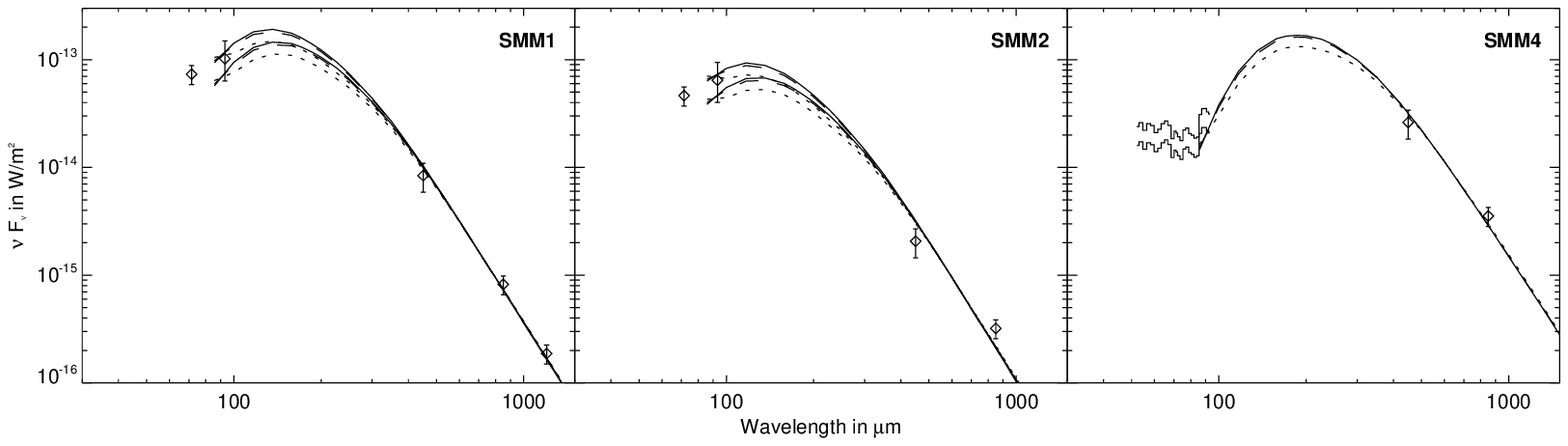}
\caption{
Spectral energy distributions of the submillimeter sources in the
region ISOSS J20298+3559.
The diamonds represent photometric data obtained with MIPS (70\,$\mu$m) and
SCUBA (450 and 850\,$\mu$m), extrapolated data at 93\,$\mu$m for SMM1 and
SMM2 and the datapoint at 1.2\,mm for SMM1.
Histogram-like bars show the error range of the MIPS SED spectrophotometry
(53-93\,$\mu$m) for SMM4.
The lines show modified blackbody fits for $\lambda \ge 93\,\mu$m using
different dust opacities (solid: no ice mantles, dotted: thick ice mantles,
dashed: thin ice mantles).
For SMM1 and SMM2 two sets of curves are shown for spectral slopes
of 0.3 and 2.0 in $70 \le \lambda \le 93\,\mu$m.
}\label{fig:20298_seds}
\end{figure*}

\begin{figure}
\centering
\resizebox{6.665cm}{!}{\includegraphics{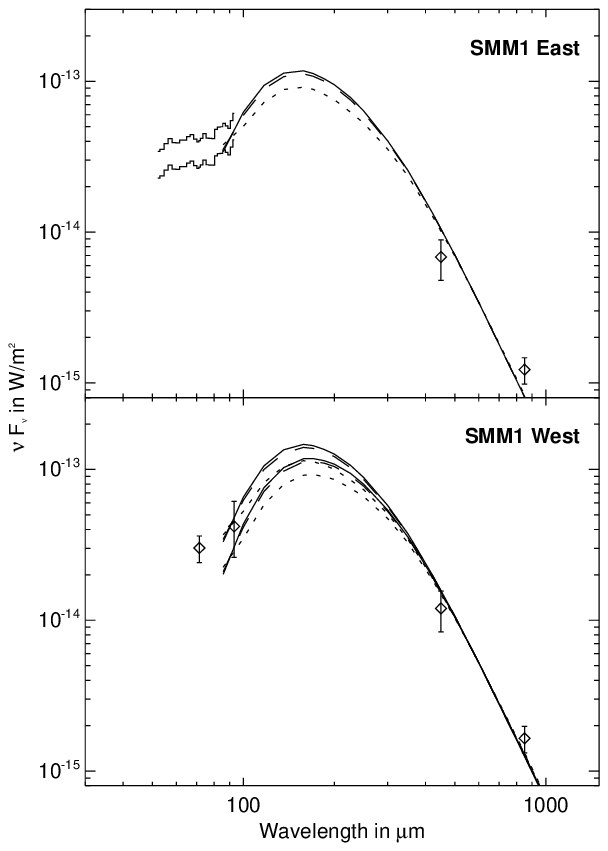}}
\caption{
Spectral energy distribution of the submillimeter sources in the
region ISOSS J22478+6357.
The diamonds represent photometric data obtained with MIPS (70\,$\mu$m) and
SCUBA (450 and 850\,$\mu$m) and extrapolated data at 93\,$\mu$m for SMM1 West.
Histogram-like bars show the error range of the MIPS SED spectrophotometry
(53-93\,$\mu$m) for SMM1 East.
The lines show modified blackbody fits for $\lambda \ge 93\,\mu$m using
different dust opacities (solid: no ice mantles, dotted: thick ice mantles,
dashed: thin ice mantles).
For SMM1 West two sets of curves are shown for spectral slopes
of 0.3 and 2.0 in $70 \le \lambda \le 93\,\mu$m.
}\label{fig:22478_seds}
\end{figure}

We first present the analysis of the long-wavelength observations
towards the star-forming regions and then address the associated infrared
sources. The extracted fluxes in the far-infrared and submillimeter are listed
in Table~\ref{tab_lw}. In
Figs.~\ref{fig:19357_seds}, \ref{fig:19486_seds}, \ref{fig:20153_sed},
\ref{fig:20298_seds} and \ref{fig:22478_seds}
the spectral energy distributions (SEDs) of the detected clumps are sketched.

\subsection{Far-infrared spectral slopes}

The examination of the SEDs compiled for the seven detected clumps towards which
MIPS SED observations have been performed shows that the spectral slopes at
wavelengths around 70\,$\mu$m and below do not resemble a single thermal emission
component that would reproduce the submillimeter fluxes. It presumably originates
from warmer dust components or transiently heated very small grains
\citep[cf.][]{2007A&A...474..883B,2005MNRAS.356..810R}. The spectral slopes
(${\rm d}\log(\lambda {\rm F_\lambda}) /{\rm d}\log\lambda$) vary in the range
$0.3 - 2.0$ for 70\,$\mu$m $<\lambda<$ 94\,$\mu$m.

\subsection{Clump dust temperatures and masses}

To characterize the cold component of dust (and gas) that gives rise to the
emission at long wavelengths, we assume that it can be reproduced by isothermal
dust emission. As noted above, the fluxes at wavelengths around 70\,$\mu$m and
below do not conform with this assumption and we therefore use the datapoints at
93, 450 and 850\,$\mu$m to estimate the dust temperature.
In that respect we also assume that the emission at
these wavelengths is optically thin and can be approximated by a modified Planck
spectrum.
We use the dust opacities given in \citet{1994A&A...291..943O} for a MRN distribution
either without ice mantles, with thin and with thick ice mantles
($\kappa_{850{\rm \mu m}}=0.6-1.4\,\rm{cm^2/g}$) corresponding to
emissivity indices of $\beta=1.81-1.85$ at long wavelengths. The dust-to-gas
mass ratio is assumed to be 1/100. In the cases where no MIPS SED measurements
are available, we extrapolate the 70\,$\mu$m flux to 93\,$\mu$m allowing for
spectral slopes of 0.3 and 2.0 (adopting the measured extremes).
We list the resulting
range of dust temperatures and gas masses for each clump in Table~\ref{tab_lw}.
The fitted curves are shown in Fig.~\ref{fig:19357_seds}, \ref{fig:19486_seds},
\ref{fig:20153_sed}, \ref{fig:20298_seds} and \ref{fig:22478_seds}.

\begin{table*}
\begin{minipage}[h]{\hsize}
\caption{Long-wavelength emission and derived properties of the cold component
of the detected clumps.}
\label{tab_lw}
\centering
\renewcommand{\footnoterule}{}  \renewcommand{\thempfootnote}{\alph{mpfootnote}}
\renewcommand{\thefootnote}{\alph{footnote}}
\begin{tabular}{c c c c c c c c c c}
\hline\hline
Region & Clump & \multicolumn{4}{c}{Total flux (Jy)} & Size (pc) & Dust temperature (K) & Gas mass (M$_\odot$) \\
ISOSS\dots & & 70\,$\mu$m & 93\,$\mu$m & 450\,$\mu$m & 850\,$\mu$m & & \\
\hline
J19357+1950 & SMM1 N & 3.3 & 6.0 & 2.1 & 0.40 & 0.37 & $16.6-18.7$ & $54-92$\\
& SMM1 S  & 0.57 & $0.80-1.3$\footnotemark[1] & 1.0 & 0.21 & 0.31 & $14.1-16.5$ & $34-66$\\
& SMM2 & 0.21 & $0.30-0.47$\footnotemark[1] & 1.9 & 0.35 & 0.39 & $12.0-13.7$ & $85-166$\\
J19486+2556 & SMM1 & $<0.15$ & $<0.26$ & 2.2 & 0.51 & 0.22 & $11.6-12.5$\footnotemark[2] & $69-123$\footnotemark[2]\\
& SMM2 & 2.7 & 4.6 & 2.1 & 0.38 & 0.14 & $16.1-18.1$ & $28-49$\\
& SMM3 & 7.4 & 11 & 4.1 & 0.59 & 0.21 & $16.8-19.0$ & $46-79$\\
J20153+3453 & SMM1 & 13 & 17 & 12 & 2.1 & 0.16 & $15.3-17.0$ & $87-149$\\
J20298+3559 & SMM1 & 1.7 & $2.5-3.7$\footnotemark[1] & 1.3\footnotemark[3] & 0.23\footnotemark[3] & 0.14\footnotemark[3] & $15.8-18.8$\footnotemark[4] & $6-13$\footnotemark[4]\\
& SMM2 & 1.1 & $1.6-2.4$\footnotemark[1] & 0.31\footnotemark[3] & 0.091\footnotemark[3] & $<0.07$\footnotemark[3] & $17.3-21.3$ & $2-3$\\
& SMM4 & 0.41 & 0.81 & 3.9 & 1.0 & 0.34 & $12.1-13.2$ & $46-80$\\
J22478+6357 & SMM1 E & 0.80 & 1.6 & 1.0 & 0.35 & 0.14 & $14.7-16.4$ & $47-81$\\
& SMM1 W & 0.72 & $1.0-1.6$\footnotemark[1] & 1.8 & 0.47 & 0.24 & $13.4-15.5$ & $79-153$\\
\hline
\end{tabular}
\footnotetext[1]{Extrapolated values.}
\footnotetext[2]{Derived using upper limit fluxes.}
\footnotetext[3]{Flux values and sizes from \citep{2003A&A...398.1007K}.}
\footnotetext[4]{Derived from listed fluxes and 0.075\,Jy at 1.2\,mm
\citep{2003A&A...398.1007K}.}
\end{minipage}
\end{table*}

\subsection{Associated infrared sources}

We detect a number of sources at 24\,$\mu$m and shorter wavelengths that are
associated with the submillimeter clumps. To evaluate their nature and
relationship with the submillimeter condensations we select the sources
marked in the mid and right panels of Fig.~\ref{fig:19357_maps} to
\ref{fig:22478_maps} for which we can identify singular counterparts. At
distances of
$2-4$\,kpc the resolution achieved in the Ks band corresponds to $1600-3200$\,AU
and we can only probe the confusion of sources on scales larger than these. For
sources that show no near-infrared counterpart the IRAC resolution at
3.6\,$\mu$m which corresponds to $3000-6000$\,AU limits the search for
multiplicity. In addition we check for optical counterparts in Second Digitized
Sky Survey DSS2-red maps. Sources without optical counterpart are best
candidates to represent embedded objects and therefore the association with the
molecular and dusty clumps is presumably no projection effect.
However, the actual positions of the sources with respect to the clumps
and the foreground extinctions are unknown.

\begin{figure*}
\centering
\includegraphics[width=18cm]{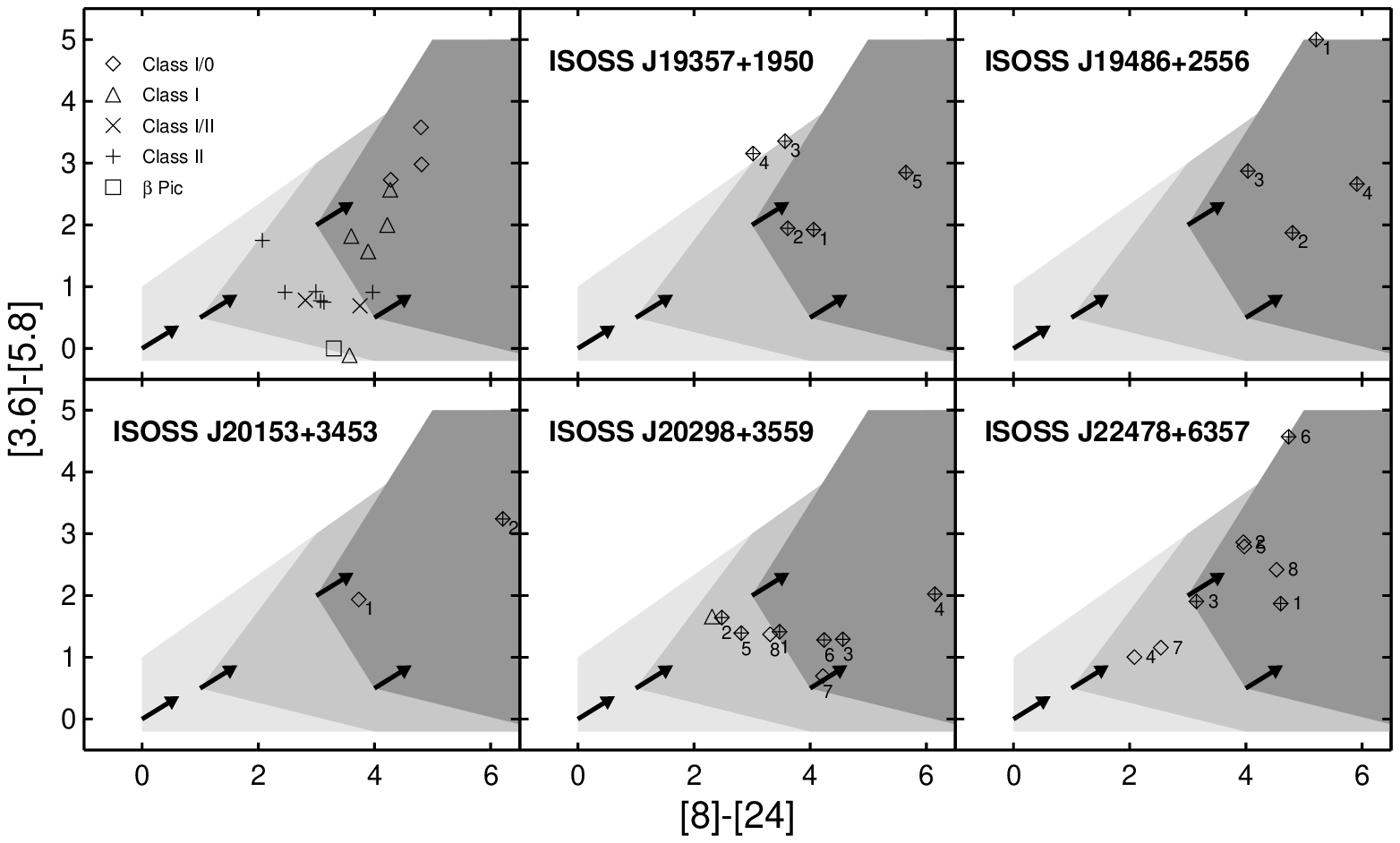}
\caption{
IRAC and MIPS infrared color-color diagram.
The regions where most models are Stage I, II and III according to
\citet{2006ApJS..167..256R} are shaded from dark to light grey.
The vectors show an extinction of A$_V$=20 \citep{2005ApJ...619..931I}.
In the upper left panel young stellar objects corresponding to different
evolutionary stages from \citet{2004ApJS..154..385R} are plotted.
In the five remaining panels the mid-infrared sources
associated with submillimeter emission in the ISOSS star-forming regions are
drawn (diamonds, numbering as in Figs.~\ref{fig:19357_maps}--\ref{fig:22478_maps}).
The sources without optical counterpart are marked by crosses.
In the panel corresponding to ISOSS J20298+3559 the approximate colors of
the Herbig B2e star are represented by the triangle.
}\label{fig:irccds}
\end{figure*}

\citet{2006ApJS..167..256R} present a grid of young stellar objects (YSOs)
model SEDs which has been used to analyze mid-infrared sources
detected with \textit{Spitzer} in star-forming regions
\citep{2007ApJ...666..321I,2007ApJ...669..464S}. Utilizing a 
Monte-Carlo radiative transfer code, more than 20\,000 2D models were computed
for ten inclinations each.
The different evolutionary phases in the modelled parameter space are
ordered according to the Class system \citep{1994coun.conf..179A} into stages:
Stage 0/I (hereafter: Stage I) are very young objects that are deeply embedded
in an extended envelope with a cavity surrounding an accretion disk, Stage II
contains disk-dominated models and Stage III represents star-dominated systems.
Via a web interface measured fluxes of individual sources can be put into a
model SED fitting routine to constrain source parameters \citep{2007ApJS..169..328R}
and we make use of this below in the Discussion section.
The color-color diagram [3.6]-[5.8] versus [8]-[24]
(Fig.~23 in \citet{2006ApJS..167..256R})
offers a good distinction between the stages. These colors have also been
compiled in \citet{2004ApJS..154..385R} for a number of young stellar objects in
the Elephant Trunk Nebula which have been classified before according to the Class
system \citep[see][and references therein]{2004ApJS..154..385R}. For a
verification of the color assessment we plot their data
and the regions corresponding to the Stages proposed by \citet{2006ApJS..167..256R}
in the upper left panel of Fig.~\ref{fig:irccds}.
The foreground extinction appears to be small for these sources.
The Class I/0 and I sources fall into the corresponding Stage I region
with one exception. The colors of Class I/II and II sources lie in the Stage II region
with one exception. The colors of the debris disk object $\beta$~Pictoris
place it on the border between the Stage II and III regions. These accordances
support the utility of the colors for classification and we apply this method to
our source samples
to get an estimate of their evolutionary status. However, due to the presence of
a significant amount of dust in the regions, an enhanced extinction is expected
for our sources.
In the remaining five panels of Fig.~\ref{fig:irccds} the infrared colors of
the sources marked (and numbered) in Fig.~\ref{fig:19357_maps} to
\ref{fig:22478_maps} are plotted. Sources without optical counterpart in DSS2-red
maps are marked with crosses. Most of the evaluated
sources exhibit colors lying in the Stage II region or in the vicinity of the
Stage I/II transition. Interestingly, a few sources show very red colors that
are consistent with Stage I even if a foreground extinction of
A$_{\rm V} \ge 100$ is assumed. Exhibiting \mbox{[8]-[24] $> 5$}, these sources are
distinct from the sample in the upper left panel.
Two red sources, Source 1 in ISOSS J19486+2556 and Source 6 in ISOSS J22478+6357,
apparently show enhanced 5.8\,$\mu$m emission but still lie within the Stage I
region. Both are very faint in the near-infrared.

Several sources are associated with ``fuzzy'' green features in the IRAC composite
images and/or extended emission features in the near-infrared:
Source 2 in ISOSS J19486+2556, Source 2 in ISOSS J20153+3453, Source 8 in ISOSS
J20298+3559 and also towards the SMM3 peak in ISOSS J19486+2556, where we cannot
identify the appropiate counterparts to the detected 24\,$\mu$m source, and
towards the SMM1 peak in ISOSS J20298+3559, where the 24\,$\mu$m source shows
no counterparts at shorter wavelengths. These features likely trace outflow
activity from young accreting objects which gives rise to shock-excitation
of molecular hydrogen and CO. However, also atomic hydrogen Br\,$\alpha$ line
emission can contribute to the IRAC 4.5\,$\mu$m flux.

The most prominent features in the IRAC composite image of ISOSS J19486+2556 are
two extended red emission features east of SMM2 and south of SMM3 which are also
visible at the MIPS wavelenghts. In the near-infrared very red objects are detected
that are located towards the centers of the features which themselves are not
traced. Extended low-density dust heated to higher temperatures could give rise
to the mid- and far-infrared emission.

In the panel of Fig.~\ref{fig:irccds} corresponding to ISOSS J20298+3559
we also plot the approximate colors of the brightest source in the region,
located western of SMM1. It has been investigated in detail in
\citet{2003A&A...398.1007K} and classified as Herbig B2e star
that is accreting from a disk which is optically thick. The 8 and 24\,$\mu$m
images seem to show non-linear effects due to the brightness of the source which
may affect the derived colors corresponding to Stage II.

\section{Discussion}

\subsection{The clump population}

Our simple approach to reproduce the long-wavelength emission towards the
sample of submillimeter clumps by one component neglects any effects of the
density and temperature distributions within the clumps and the assumption of
optically thin
emission may be violated in the far-infrared if the (column) density is high
enough. Assuming a density profile that is flat in the innermost region of
100\,AU radius, follows a power-law (${\rm \rho \propto r^{-\alpha}}$
with $\alpha$ = 1.5) in between and is cut-off exponentially at 20\,000\,AU,
an optical depth of unity is reached
at about 100\,$\mu$m for a 100\,M$_\odot$ condensation (dust opacities from
\citet{1984ApJ...285...89D}; Steinacker, private communication). Therefore
we may underestimate the total emission at the shortest wavelengths used
(93\,$\mu$m) and also the mean dust temperatures. Setting
the dust temperature to a canonical value of 20\,K results in a decrease of the
mass by up to a factor of about two for the
coldest temperatures, but would require high optical depths at wavelengths
around 100\,$\mu$m to be consistent with the measured fluxes.
However, the required mass scales similarly to the optical depth.
 It seems therefore
unlikely that this effect causes errors in the mass estimation of more than
a factor of two.

The distances to the regions are mainly derived from radial velocities and
carry therefore some uncertainty. In \citet{2003A&A...398.1007K} the distance to
ISOSS J20298+3559 has been constrained by an extinction study to
$1.8 \pm 0.3$\,kpc. We
estimate the other distances to be reliable within $\sim 1$\,kpc which
results in possible mass changes by factors of 1.3 to 1.8.
Another uncertainty factor of the order of two for the mass estimate arises
from the unknown dust
composition, e.g. the dust opacities given by \citet{1994A&A...291..943O} are
about five times higher than those used for the calculation
of the model SED grid \citep{2003ApJ...591.1049W}.
It is obvious from the SED diagrams that the used dust models in combination with
the one-component approach does not fit the data well for all sources: In the
cases of SMM2 in J20298+3559 and SMM1 East in J22478+6357 it seems that
dust models corresponding to a lower value of the emissivity index $\beta$ at
long wavelengths or the introduction of several emission components would
improve the fit. However, the latter approach would increase the number of
variable parameters for the fit above the number of datapoints used. From the
available measurements in the submillimeter range we cannot place constraints on
the dust properties.

The derived dust temperatures (see Table~\ref{tab_lw}) lie between 11.6 and
21.3\,K, approximately in the same range
as the far-infrared color temperatures from ISO and IRAS fluxes of the whole
regions (see Table~\ref{tab_regions}).
In comparison to other studies, the values are consistent with those of cold dust
components derived and assumed for young dense clumps
\citep[e.g.][]{2007ApJ...656L..85B,2005ApJ...634L..57S,2007ApJ...666..982E}.

The estimated masses that reside in these cold dust and gas components range
from 2\,M$_\odot$ up to 166\,M$_\odot$.
Taking the mass estimates at face value, four of the clumps may be massive
enough ($> 100$\,M$_\odot$) to form a high-mass star and an accompanying
cluster, if one assumes that a fraction of 10\% of the total mass ends up in
massive stars \citep{Zinnecker2007}. The star formation efficiency is
observationally not well constrained on the scales of individual clumps.

From the source sizes in Table~\ref{tab_lw} volume-averaged densities
have been derived for our clump sample under the assumption of spherical symmetry.
The resulting densities are ${\rm 4\cdot10^4\,cm^{-3} < n_{H_2} < 10^6\,cm^{-3}}$ with
a median value of ${\rm 2\cdot10^5\,cm^{-3}}$.
Compared to studies of dense condensations found in IRDCs
\citep[][median density ${\rm 3\cdot10^4\,cm^{-3}}$]{2006ApJ...641..389R}
and in Cygnus~X
\citep[][median density ${\rm 2\cdot10^5\,cm^{-3}}$]{2007A&A...476.1243M}
the evaluated clumps seem to be more similar to the latter sample based on a
millimeter continuum survey.
Both the mentioned samples seem to contain candidate objects
representing very early stages of massive star formation
(e.g. massive infrared-quiet cores).
For the sample of \citet{2002ApJ...566..945B} based on IRAS measurements and
containing luminous infrared sources a lower mean density of
${\rm 9\cdot10^3\,cm^{-3}}$ is recalculated in \citet{2007A&A...476.1243M}.
The direct comparison of these densities may be affected by systematically
different derivation methods, but it suggests that the source selection
based on longer wavelengths gives access to presumably less evolved sources.
In the present case the 170\,$\mu$m band measurements constrain the average dust
temperature to lower values than those that can be inferred from the IRAS bands.

\subsection{The nature of the 24\,$\mu$m sources}

The majority of the clumps appear in a stage where 24\,$\mu$m sources are present.
In the following we discuss their ability to generate high-mass star progenitors
and evaluate selected subsamples in more detail.
During their formation high-mass stars develop high luminosities: According to
the approach of \citet{2003ApJ...585..850M} values above
100\,L$_\odot$ are reached already when one solar mass has been accreted.
The corresponding timescale is on the order of several 10$^4$\,years.
Therefore young high-mass stars should stand out as luminous sources.
However, due to the reprocessing of the radiation at short wavelengths by the
surrounding dusty envelopes only a fraction of the luminosity is released
at the infrared wavelengths we probe (here we consider $\lambda \le 24$\,$\mu$m).
If the source is deeply
embedded, we will expect its infrared colors to be very red and the emission at
optical wavelengths to be not detectable. By combining the classification from
the colors with an estimate of the total luminosity including the far-infrared
and submillimeter data we therefore get a handle on the
probability of a source to be a high-mass star precursor: Sources of Stage II
or later should generate high luminosities to be considered candidate massive
young stellar objects.

We consider the sources' locations
in the infrared color-color space as qualitatively describing their evolutionary
stages. From the sample of 24\,$\mu$m sources that has been evaluated in the Analysis
section the majority is estimated to be comparable to Stage II models. As the
majority of the associated clumps does not substantially exceed a
total luminosity of 200\,L$_\odot$, we consider most of these objects as being
rather low-mass YSOs. Their association with the submillimeter clumps suggests
that they have formed inside and that the clumps contain density substructures.

\subsubsection{Very red protostar candidates}

As described above, sources that display very red infrared colors and sources
that are not detected in the shorter-wavelength bands are best candidates to
represent very young objects that may evolve to massive YSOs. Using
Fig.~\ref{fig:irccds} we select Source 5 in ISOSS J19357+1950, Source 1 in ISOSS
J19486+2556 and Source 2 in ISOSS J20153+3453. The remaining sources with
\mbox{[8]-[24] $> 5$} are not associated with massive submillimeter clumps and
probably will not be able to accrete enough material to end up as high-mass
stars. Furthermore, in \citet{2003A&A...398.1007K} the submillimeter source SMM1
in ISOSS J20298+3559 is found to be a candidate Class 0 object. We detect a
24\,$\mu$m source towards this peak which is not seen at shorter wavelengths.
Associated outflow-tracing emission supports the classification as young
protostellar object.

Source 5 in ISOSS J19357+1950 is located towards the massive clump SMM2. We
apply the online model fitter using the
fluxes including upper limits in the H band, at 70, 450 and 850\,$\mu$m.
Source 2 found towards the massive clump ISOSS J20153+3453 SMM1
exhibits very red colors and is associated with a possible outflow.
The online model fitter is fed with the fluxes including upper limits in the
J band, at 70, 450 and 850\,$\mu$m.
For SMM1 in ISOSS J20298+3559 we input the fluxes measured at 24, 70, 450 and
850\,$\mu$m and upper limits in the IRAC bands and in the Ks band.
The range of possible foreground extinction is selected to ${\rm A_V \le 500}$.
The fitting results for these three sources are listed in Table~\ref{tab_ysos1}.
From the inspection of the SED plots we establish a cut-off value of
$\chi^2_{\rm d}$ ($\chi^2$ per datapoint) to select the models that fit the
datapoints within reasonable margins.
We also give the total extinction to the central source by combining the fitted
foreground extinction with the extinction along the line of sight from the
model outer radii inwards.
These parameters are degenerate to some extent and allow models with
different envelope sizes to be fitted via compensating foreground extinctions.
 The results are consistent with the interpretation
of the sources as young embedded stellar precursors.
The observed outflow activity signatures accord with high accretion rates for
the latter two sources. In the cases of ISOSS J19357+1950 SMM2 and
ISOSS J20153+3453 SMM1 the estimated large total masses of around
100\,M$_\odot$ residing in cold components of dust and gas would in principle
allow ongoing accretion and the build-up of high-mass stars.
However, due to the multiplicity of
near-infrared sources that are detected in the vicinity of ISOSS 20153+3453 SMM1
the further evolution of this object is not straightforward. Towards
ISOSS J19357+1950 SMM2 there also is a second more evolved source nearby.
In the third case the mass estimate of about 10\,M$_\odot$ for
ISOSS J20298+3559 SMM1 is consistent with the source being an intermediate-mass
star precursor.

The remaining very red candidate Stage I source located towards the clump SMM1
in the region ISOSS J19486+2556 is particularly interesting
because of the low infrared luminosity in comparison to the submillimeter,
making it a candidate for a young and deeply embedded object.
We input the photometric data into the online model fitter including upper
limits in the Ks band, at 70, 450 and 850\,$\mu$m. We choose a cut-off at
$\chi^2_{\rm d} < 49.6$ to select the acceptable fits. The results
for this source are ambiguous: Among the selected models are several belonging
to the Stage I category with central source masses of about 0.1\,M$_\odot$ but
also a majority of Stage II models with central source masses between 3 and
4.5\,M$_\odot$, both with and without envelopes. There is no significant
difference in the fitted total extinction towards the central sources. We
therefore cannot assess and constrain the parameters of this source.

\begin{table*}
\begin{minipage}[h]{\hsize}
\caption{Results of the SED fitting for three very red young stellar objects.}
\label{tab_ysos1}
\centering
\renewcommand{\footnoterule}{}  \renewcommand{\thempfootnote}{\alph{mpfootnote}}
\renewcommand{\thefootnote}{\alph{footnote}}
\begin{tabular}{l c c c}
\hline\hline
Region ISOSS\dots & J19357+1950 & J20153+3453 & J20298+3559\\
Source & 5 & 2 & -\\
Associated clump & SMM2 & SMM1 & SMM1\\
\hline
Selected cut-off & $\chi^2_{\rm d} < 14.2$ & $\chi^2_{\rm d} < 9.3$ & $\chi^2_{\rm d}<12.5$\\
Fitted models & Stage I & Stage I & Stage I\\
Total extinction & ${\rm A_V\sim85}$ & ${\rm A_V\sim100}$ & ${\rm A_V > 200}$\\
Central mass (M$_\odot$) & $0.2-2.1$ & $0.7-2.3$ & $0.2-4.6$\\
Envelope accretion & & &\\
\quad rate (M$_\odot$/yr) & $1.6\cdot10^{-6}-8.9\cdot10^{-5}$ & $1.5\cdot10^{-6}-3.4\cdot10^{-4}$ & $4\cdot10^{-6}-7\cdot10^{-4}$\\
\hline
\end{tabular}
\end{minipage}
\end{table*}

\subsubsection{Evolved young stellar objects}

We also detect sources whose SEDs are dominated by emission at mid-infrared
wavelengths which exhibit higher total luminosities and may represent
more massive YSOs. By comparing the spectral energies at 24\,$\mu$m and
70\,$\mu$m we find two candidate sources, located towards SMM4 in
ISOSS J20298+3559 (Source 8) and SMM1 East in ISOSS J22478+6357 (Source 3).
We also use the fluxes from \citet[IRS 6]{2003A&A...398.1007K} for the I, J, H
and Ks bands as well as upper limits at 70, 450 and 850\,$\mu$m for the former
and upper limits in the J band, at 70, 450 and 850\,$\mu$m for the latter source.
The fitting results are compiled in Table~\ref{tab_ysos2}. We consider the first
source to be an embedded proto- or pre-main-sequence
star of intermediate mass as also suggested by \citet{2003A&A...398.1007K}.
It probably has been formed in the SMM4 clump. The results for the second source
are consistent with the source representing an evolved YSO of
at least intermediate mass that is still embedded and presumably has been formed
within the clump SMM1 East.

\begin{table}
\begin{minipage}[h]{\columnwidth}
\caption{Results of the SED fitting for two evolved young stellar objects.}
\label{tab_ysos2}
\centering
\renewcommand{\footnoterule}{}  \renewcommand{\thempfootnote}{\alph{mpfootnote}}
\renewcommand{\thefootnote}{\alph{footnote}}
\begin{tabular}{l c c}
\hline\hline
Region ISOSS\dots & J20298+3559 & J22478+6357\\
Source & 8 & 3\\
Associated clump & SMM4 & SMM1 East\\
\hline
Selected cut-off & $\chi^2_{\rm d}<3.3$ & $\chi^2_{\rm d}<2.1$\\
Fitted models & Stage II\footnotemark[1] & Stage II\footnotemark[2]\\
Total extinction & ${\rm 6<A_V<9.4}$ & ${\rm 29<A_V < 41}$\\
Total luminosity (L$_\odot$) & $\sim 150$ & $\sim 2000$\\
Central mass (M$_\odot$) & $3.2-4.9$ & $6-8.5$\\
Preferred inclination & low & -\\
Disk mass (M$_\odot$) & $4\cdot10^{-6}-10^{-1}$ & $10^{-5}-10^{-1}$\\
Accretion rate (M$_\odot$/yr) & $10^{-11}-10^{-6}$ & $5\cdot10^{-11}-2\cdot10^{-5}$\\
System age (yr) & $10^6-10^7$ & $10^6 - 6\cdot10^6$\\
\hline
\end{tabular}
\footnotetext[1]{M$_{\rm disk}$/M$_*\ge7.7\cdot10^{-7}$}
\footnotetext[2]{M$_{\rm disk}$/M$_*\ge9.5\cdot10^{-7}$}
\end{minipage}
\end{table}

\section{Conclusions}

We have analysed multi-wavelength observations of five star-forming regions
that were identified using the ISOPHOT Serendipity Survey at 170\,$\mu$m.
From the discussed results we infer:
\begin{enumerate}

\item We found one to four compact ($\sim 0.2$\,pc) submillimeter
condensations in
every region that represent molecular clumps containing a cold component
of gas and dust. The dust temperature estimates vary between 11.6 and
21.3\,K and accord with the large-scale color temperatures measured in the
far-infrared.

\item The resulting estimated clump masses that reside in these cold components
range from 2 to 166\,M$_\odot$. Four out of twelve clumps may be massive enough
($> 100$\,M$_\odot$) to be promising candidate birthplaces for high-mass stars.

\item We identify multiple associated mid-infrared sources for the majority
of the clumps suggesting that they embody significant density substructures
on smaller scales than probed by the submillimeter observations ($\gtrapprox 0.1$\,pc).
Since the emission of most clumps is dominated by
the cold material we expect thriving star formation from further collapse
of overdensities.

\item Most of the associated sources are considered as low-mass young stellar
objects in an evolutionary state later than Class 0/I partly embedded in the
clumps where they have formed.

\item In the region ISOSS J19357+1950 we find a massive clump of around
$100$\,M$_\odot$. A probably deeply embedded accreting Class 0/I protostar of
$\sim 1$\,M$_\odot$ is detected adjacent to a more evolved young stellar object.
Similarly, in ISOSS J20153+3453 one massive clump of about $100$\,M$_\odot$
is present and a cluster
of sources is detected in the vicinity containing a presumably deeply embedded
accreting protostar with a current mass of $\sim 1.5$\,M$_\odot$. Its
evolutionary state corresponds to Class 0/I and there is evidence
for outflow activity from this object.

\item The source SMM1 in ISOSS J20298+3559 probably represents a young
accreting intermediate-mass star precursor embedded in a molecular clump of
$\sim 10$\,M$_\odot$. Towards SMM4 in the same region we identify an embedded
proto- or pre-main-sequence star of $3.2-4.9$\,M$_\odot$ that likely evolved
from the associated clump of $\sim 60$\,M$_\odot$.

\item In the region ISOSS J22478+6357 a candidate protostar
of $6-8.5$\,M$_\odot$ is found that is embedded in the associated
SMM1 East clump ($\sim 60$\,M$_\odot$).

\item We do not detect stellar precursors that could have current masses
of 10\,M$_\odot$ or more. Therefore we cannot be sure whether such objects
will emerge within the identified clumps. However, the presence of
intermediate-mass proto- or pre-main-sequence star candidates and the
large clump masses indicate that one can expect that also high-mass stars are formed
in these systems.

\item Our study reveals that the search for regions containing large fractions
of cold material in the far-infrared has successfully identified star-forming
regions associated with cold and massive clumps. Compared to sources selected
at shorter wavelengths they may represent less evolved stages of high-mass
star-formation and in that respect substantial impact can be expected from
future (space) missions like Herschel and Planck. Furthermore this study
demonstrates that the detailed star-forming content can only be accessed with a
multi-wavelength approach including sensitive mid- and far-infrared observations.

\end{enumerate}

\begin{acknowledgements}
Based on observations with ISO, an ESA project with instruments funded by ESA
Member States (especially the PI countries: France, Germany, the Netherlands and
the United Kingdom) and with the participation of ISAS and NASA. The ISOSS was
supported by funds from the DLR, Bonn.
Based on observations collected at the
Centro Astron\' omico Hispano Alem\' an (CAHA) at Calar Alto, operated jointly
by the Max Planck Institut f\"ur Astronomie and the Instituto de Astrof\' isica
de Andaluc\' ia (CSIC), and on observations with the
James-Clerk-Maxwell Telescope (JCMT) as well as on
observations made with the Spitzer Space Telescope, which is operated by the Jet
Propulsion Laboratory, California Institute of Technology under a contract with
NASA.
This publication makes use of data products from the Two Micron All Sky Survey,
which is a joint project of the University of Massachusetts and the Infrared
Processing and Analysis Center/California Institute of Technology, funded by the
National Aeronautics and Space Administration and the National Science
Foundation.
The Digitized Sky Surveys were produced at the Space Telescope Science Institute
under U.S. Government grant NAG W-2166.
The Second Palomar Observatory Sky Survey (POSS-II) was made by the California
Institute of Technology with funds from the National Science Foundation, the
National Geographic Society, the Sloan Foundation, the Samuel Oschin Foundation,
and the Eastman Kodak Corporation.
MH thanks J\"urgen Steinacker and Ulrich Klaas (both at MPIA) for their support
and helpful discussions; Kalevi Mattila and Jos\'e Gon\c{c}alves
(University of Helsinki Observatory) for hospitality and fruitful comments.
\end{acknowledgements}

\bibliographystyle{aa}
\bibliography{ms}

\end{document}